# Negative Compressibility of Single Selenium Chain Confined in Zeolite Pore

Wei Ren[1], Jian-Ting Ye[1,2], Wu Shi[1], Zi-Kang Tang[1], C. T. Chan[1], and Ping Sheng[1]

[1] Department of Physics, Hong Kong University of Science and Technology

Clear Water Bay, Kowloon, Hong Kong, China

[2] Institute for Material Research, Tohoku University, Katahira 2-1-1, Aoba-ku, Sendai 980-8577, Japan

**Pressure induced structural and electronic transitions of Se helical chains confined inside nano-channels are studied. Raman scattering and optical absorption experiments show strong evidence of band gap reduction under high pressure. *Ab initio* calculations reveal that under hydrostatic compression, the Se chains should elongate and the change in morphology leads to a softening of phonons and narrowing of band gaps, and these signatures are observed in experiments. Our investigation demonstrates a negative compressibility in one dimension.**

Selenium is a group VI non-metallic element with a band gap of about 2 eV at ambient pressure. To realize metallization for bulk selenium with a concomitant structural phase transition, a large pressure up to tens of GPa is necessary[1-3]. Such phase transition phenomena of polycrystalline selenium have been observed at very high pressures by in situ Raman and X-ray diffraction. The metallization[4] and anomalous liner expansion coefficient under 20 GPa pressure[5] have been examined for crystalline selenium using density functional calculations. Here, we investigate the property of a single Se chain under pressure and we demonstrate that a much lower pressure is sufficient to induce detectable changes in the properties of a confined Se chains inside nano-channels. The pressure induced structural and electronic changes of Selenium chains confined in zeolite channels are studied using Raman spectroscopy and optical absorption spectra, and the results hint at a trend towards metallization as we increase the pressure. Such changes are interpreted using density functional theory (DFT) calculations. The calculations focus on understanding the hydrostatic pressures effect applied on the single trigonal helical chain structures. The first-principles calculation results are in excellent agreement with the experimental findings.

Our zeolite template AlPO$_4$-5 crystals (IUPAC code AFI; molecular formula Al$_{12}$P$_{12}$O$_{48}$) are synthesized using a hydrothermal method[6]. The crystalline framework with parallel one-dimensional channels is formed by alternating tetrahedra of [AlO$_4$]$^-$ and [PO$_4$]$^+$ units (space group: P6/mcc). The inner diameter of the channel is 7.3 Å, and the separation distance between two neighboring channels is 13.7 Å. The selenium species are incorporated into the AlPO$_4$-5 crystal channels by vapor phase diffusion method as described previously[7-9]. Earlier X-ray absorption spectroscopy study on the Se-Se distances has confirmed that single chains of Se are indeed being confined in the assembling template[10].

Visible light Raman spectra of our isolated selenium chains are shown in figure 1. The response of a pure Zeolite sample is much weaker than that with selenium inside its channels, and thus can be neglected. We have previously studied the six optical phonon modes $\Gamma_{(optic)}$=A$_1$+A$_2$+2E$_1$+2E$_2$, shown partially in figure 1. The highest peak (257cm$^{-1}$) with single frequency A$_1$ is the symmetric chain radial expansion mode, rather similar to radial breathing mode in carbon nanotubes. The mode A$_2$ is chain rotation mode and Raman-inactive. Both E modes are doubly-degenerate, while one of them is peaked at 233 cm$^{-1}$ and the other is well below our present Raman frequency. These modes have ZZ configuration, in which both the excitation and the scattering light are parallel to the chain direction. We can also identify a peak for molecular Se at 267cm$^{-1}$, which appear in both ZZ and YY (both excitation and the scattering light are perpendicular to the chain axis) configurations. This peak is attributed to the crown-shaped Se8 ring structure[11, 12]. As we increase the pressure, the most significant change is that the low-frequency tail rises up while the high-frequency signal suppresses down. Such kind of red-shift of oscillator strength has been discussed before by others on the basis of chain-chain interaction[13]. There is no inter-chain coupling in our structure, and thus our observed pressure-induced frequency magnitude change has a different origin. As pressure up to 6.42 GPa is being applied in the channels through a pressure-transmitting liquid solution (ethanol to methanol 1:4 in volume percentage), we may think that the observed changes are due to a longitudinal compression of the individual Se chains. Quite on the contrary, we found the surprising result that the Se chains are elongated, as we will establish in the following discussions. In short, both the Se-Se bond length and angle are increased by the pressure. This structural change gives rise to the softening of vibration modes.

As mentioned, cyclic Se molecules such as rhombohedral Se6 or $\alpha$-monoclinic Se8 rings could also exist in the channels, since the defects in the zeolite crystal might possibly accommodate these stable molecular structures. Electronically the ring clusters show a similar semiconducting band gap as the trigonal chain. Bulk Se chain and ring structures under high pressure have also been investigated by X-ray and Raman spectroscopy measurements. The Raman frequencies showed a similar pressure-induced softening[14]. However we want to emphasize again that previously it was interpreted in terms of an interference effect between inter- and intra-molecular bonds in the molecular crystals[13]. In the bulk material, the intra-chain Se-Se covalent bonds are much stronger than the inter-chain Van der Waals interactions. The applied pressure would pack the chains denser and thus increase chain-chain coupling rapidly. In our experiment, we have eliminated this mechanism since we are dealing with single selenium spiral chains and inter-chain coupling does not enter into the picture here.

An isolated single trigonal Se helical chain is energetically more favorable compared with a linear or zigzag chain[15]. This reflects the fact that bulk Se crystal can be viewed as an array of such helices with weak inter-chain couplings. However, a trigonal Se helix is a semiconductor while both of the hypothetical linear or zigzag chains are metallic (coordination number equals 2 in all cases).

To model the hydrostatic pressure applied on the Se chain in the confined channels, we have studied various compressed and decompressed chains. We do not consider Al, P and O atoms on the wall of AFI channels, simply assuming that Se chains are unaffected by weak interaction with these host atoms. Furthermore, the incommensurate periodicity and huge computation expense definitely prevent us from including the channel template material. In our simulation the compression and decompression are done along the chain axis, while the symmetry remains unchanged. Interestingly the compressed chains give no significant band gap modification. On the other hand, the decompressed chains turn out to have gradually narrowing band gaps. What we observed experimentally is that band gap indeed decreases as the applied pressure increases. Figure 2 shows the optical absorption peaks along different polarizations of the incident light. The polarized optical absorption spectroscopy shows a marked anisotropy, where E||c (polarized along the

channel direction) absorption is generally much higher than E⊥c (polarized perpendicular to the channel direction). Such anisotropy is the typical absorption spectra of elongated objects, for if the objects are in form of clusters or rings, the dependence on polarization should be weak. This thus provides strong evidence that the Se atoms are in the form of long chains inside the channels.

As we increase the pressure, the absorption shifts conspicuously to the lower energy region for E‖c, while there is a very small shift towards higher frequency for E⊥c. These facts provide additional evidence of existence of Se chains with regularity in the AFI channels. Once again the band gap suppression is due to the same reason of chain lengthening. Electronic band structure calculations indicate that the longer the chains, the narrower the band gaps. Inset of figure 2a shows the band structure under local density approximation (LDA) of the Se chain for zero pressure (with $k$ along the chain axis, 3 atoms per unit cell). Selenium atom has the electronic configuration of $4s^2 4p^4$. As shown in the inset, two 4p electrons contribute to form covalent $\sigma$ bonds with two neighboring atoms. The remaining two 4p electrons form the lone pair (LP) states and the highest valence bands. Above the Fermi level is the empty 4p anti-bonding $\sigma*$ states for the lowest unoccupied conduction bands. Our calculations confirmed this gap shrinkage when the chain is lengthened by 0.1 Å, and thinned by 5%. Therefore the red shift of absorption threshold is corresponding to pressure induced negative compressibility along the chain direction. Upon release of pressure, we have observed the absorption spectra to be recovered (not shown here).

This result suggests that the trigonal chain's length increases counter-intuitively with increasing the hydrostatic pressure. We are able to understand this effect in the following discussions. Trigonal chalcogen phases (Se and Te) are known to have unusual negative linear compressibility[16] among some other rare materials. That means the spiral selenium chain diameter contracts and the chain length expands under pressure. Figure 3a shows a total-energy map of Se chain as a function of helical radius r and length lattice constant c. Using DFT method, we first obtain the equilibrium structure of the helical Se chain located in the center of the plot. Contraction or expansion in all three dimensions gives very high energies in the landscape, and so the anisotropic deformation is favorable under pressure. Away from equilibrium, the Se chain can minimize deformation energy either by elongating in length and at the same time contracting in radius, or it can shorten in length and expand in radius simultaneously. We will show in the following that the chain will choose to elongate in length and shrink in radius in order to be compatible with the hydrostatic external pressure. Once the structural parameters are different from their equilibrium values, there are finite energy gradients along structural coordinates (i.e. $\frac{\partial E}{\partial r} \neq 0$ and $\frac{\partial E}{\partial c} \neq 0$), and when properly normalized by an area, these gradients correspond to a pressure. If we want to deform the Se chain to the non-equilibrium position, a corresponding external pressure tensor has to be applied to the chain. As the experiment applies hydrostatic pressure, the boundary condition is to have equal pressure along radial and longitudinal directions $P(V) = -\frac{\partial E(V)}{\partial V} = -\frac{1}{2\pi r_0 c_0} \frac{\partial E}{\partial r} = -\frac{1}{\pi r_0^2} \frac{\partial E}{\partial c}$, where $r_0 = 0.976 \, \text{Å}$ and $c_0 = 4.962 \, \text{Å}$ are the chain radius and length under zero pressure.

The quantities $-\dfrac{1}{2\pi r_0 c_0}\dfrac{\partial E}{\partial r}$ and $-\dfrac{1}{\pi r_0^2}\dfrac{\partial E}{\partial c}$ can be regarded as the pressure along the radial and tangential direction that would be needed to maintain the chain deformation. The pressure vector components $\left(-\dfrac{1}{2\pi r_0 c_0}\dfrac{\partial E}{\partial r}, -\dfrac{1}{\pi r_0^2}\dfrac{\partial E}{\partial c}\right)$ are illustrated in figure 3b as a vector map. The arrows of 45 degrees give the unidirectional pressure line and their magnitudes determine the hydrostatic pressure applied.

We note that there are four quadrants in Fig. 3a and 3b, corresponding to the four possible combinations of $(\pm\delta r, \pm\delta c)$ about the equilibrium point at $(\delta r, \delta c) = 0$ for a freely standing spiral chain. If we impose any external condition of equal pressure on the radial and tangential direction, the chain will be changed from the equilibrium condition of $\dfrac{\partial E}{\partial r} = \dfrac{\partial E}{\partial c} = 0$ to a compressed configuration such that $-\dfrac{1}{2\pi r_0 c_0}\dfrac{\partial E}{\partial r} = -\dfrac{1}{\pi r_0^2}\dfrac{\partial E}{\partial c}$. We found that such condition can only be found in the quadrant $\delta r < 0, \delta c > 0$, corresponding to configurations in which the spiral chain become "slimmer" and "elongated". The configuration corresponding to the pressure of 6 GPa (the experimental nominal pressure is up to 6.42 GPa) is found to be $(\delta r, \delta c) = (-0.0488, 0.1)$ measured in Angstrom.

In the following, we examine the change in electronic and lattice vibrational properties as the chains are elongated under pressure. We also show the LDA band gaps of the Se chains for different values of $(\delta r, \delta c)$ for comparison. The contour plot shows clearly that deformations in the second quadrant give reduced band gap. In the extreme case, the top left corner shows a metallization with zero band gap, corresponding to exactly $\delta r < 0, \delta c > 0$ quadrant which we identified. Interestingly, the isotropic contractions $\delta r < 0, \delta c < 0$ under pressure would lead to a dramatically enlarged band gap. This common sense pressure response cannot explain what happens when selenium chains are compressed hydrostatically. Instead it transforms the Se chains to an insulator which is against to our experimental observations.

We now discuss the effect of elongation on phonon frequencies. We expect that when the chains are elongated, the weakening of neighboring atomic interaction should lead to reduced phonon frequencies. Such softening effects of tensile strain response are well known in quasi-one-dimensional systems such as carbon nanotubes[17]. Our numerical results indeed found that helical Se chains behave in a similar manner. Figure 4 shows the calculated phonon frequencies as a function of $\delta c$. For each value of $\delta c$, we relax the atoms to their zero force position and we compute the zone-center phonon frequencies by computing and diagonalizing the force matrix. We also computed the phonon frequencies of an isolated Se8 atomic ring (fully relaxed, see inset for atomic geometry) and results are plotted in Fig. 4 for comparison. We first note that the highest two Raman active modes of the equilibrium $(\delta c = 0)$ Se chain are slightly below the highest frequency mode of the 8-atom Se ring. This qualitatively compares very well with the experimental results shown in Fig. 1c. When the $\delta c$ is changed, there is a decrease (increase) of phonon frequencies as the chain is elongated (shortened). We note that in Fig. 1b for the measured Raman spectra under high pressure, there is an overall reduction of signal intensity, but the signal strength is skewed towards the low frequency side of the spectra. We cannot say for sure whether it is a signature of the softening of the Se chain, but it is perhaps fair to say that the results are not inconsistent with the LDA predicted softening for $\delta c > 0$.

The volume compression actually results in contraction of chain radius, longer Se-Se bond length and larger bond angle. This extraordinary picture makes all the experimental and theoretical data fully consistent. Brief mention should be made of the marked directional negative compressibility, which might not be strange in a class of compounds and biological systems[18]. But to the best of our knowledge the single element Se chains response to pressure has not been studied.

In conclusion, we investigated the properties of Se atomic chains under pressure inside one-dimensional nano-channels. We found that inside the nano-channels, just a moderate external pressure can induce quite conspicuous change in structural, electronic, optical and vibrational properties. The confined geometry allows us to investigate the effect of pressure on a single chain. Density functional calculations show that the Se chain should be elongated under pressure, and as a consequence, the band gap should narrow and phonon frequencies should become softer, as observed experimentally. We presented a means of quantum control realized by the application of hydrostatic pressure. This remarkable property may find applications in the deep ocean telecom systems and sensitive pressure sensors.

**Methods**

Measurement details:

Optical characterizations of selenium nanostructures formed inside $AlPO_4$-5 single crystals were measured by a Jobin-Yvon T64000 Raman spectrometer equipped with a CCD detector cooled by liquid nitrogen. The Raman scattering was excited by the 632.8 nm line from a He-Ne laser. A 100W xenon lamp was used as the light source for optical absorption spectroscope. Using the confocal technique, optical absorption spectroscopy were measured within a light spot of less than 20 microns in diameter with its center coincide with the 1 μm light spot of the excitation laser for Raman scattering. The hydrostatic pressure was applied to the crystal by using a diamond anvil cell (DAC). The crystal was immersed in the transmitting medium ethanol and methanol mixture together with a ruby grain which is used to determine the pressure inside the stainless steel gasket. The DAC is pressed *in. situ.* by a gas membrane without moving the sample.

 Computational details:

The Vienna *ab initio* simulations program[19] (VASP4.6.25) was used to perform the first-principle calculations. An energy cutoff of 194.0 eV (14.26 Ry) / Perdew-Wang (PW91), and cutoff 264.4 eV (19.43 Ry) / Perdew–Burke–Ernzerhof (PBE) exchange-correlation functional were applied. Infinite selenium trigonal chains and finite selenium rings are simulated. The Brillouin zone was sampled with 1x1x30 *k* points of a Monkhorst-Pack grid for chains. The supercell geometry is used and the chain-chain distance is chosen as 10A. At such as distance, interactions between adjacent chains are very small so that the results can be interpreted as those for a single chain. Before the phonon calculation each Se chain structure of fixed repeat length was fully relaxed with respect to the other two dimensions according to calculated force on the atoms. The Hellmann-Feynman forces are converged to less than 0.005 eV/A.

 **Acknowledgement**

We are grateful to Prof. Graeme J. Ackland and Prof. T.C. Leung for helpful discussions. We gratefully acknowledge financial support of RGC CERG HKUST 602807, DAG05/06.SC33, (Z.K.T.), RPC06/07.SC21 (C.T.C) and CA04/05.SC02 (P.S.). The computation resources were supported by the Shun Hing Education and Charity Fund.

Correspondence and Request for Materials should be addressed to Z.K.T (phzktang@ust.hk) and P.S. (sheng@ust.hk).

**Author contributions**

J.T.Y and W.S. carried out the sample preparation and measurement experiments. W.R. and C.T.C. performed the calculation and theoretical work.  Z.K.T and P.S. designed and guided the whole work. W.R. and C.T.C wrote the paper.

**Figure captions:**

**Figure 1** (a) Schematic representation of Se helical chains in the AlPO$_4$-5 single crystal framework. Raman spectra of the trigonal selenium chain plotted as a function of applied pressure. As the pressure increases (b), Raman mode softening is gradually observed where low frequency spectra dominate. When we decreased the pressure (c), the Raman spectra were recovered back. Left and right insets in (c) show the ZZ and YY polarized Raman signals.

**Figure 2** Optical absorption of selenium chain measured in a high-pressure diamond anvil cell at 0 to 6.42 GPa. The onset shifts of parallel polarized optical absorption (a) indicate the suppression of selenium band gap. Inset shows our calculated LDA band structure where Fermi energy is equal to 0 eV. (b) For the perpendicular polarized optical absorption, an unnoticeably small peak shift to higher photon energy.

**Figure 3** (a) Total energy contour plot is shown with varying cross sectional radius and unit cell length along chain axis. (b) The gradient of total energy plot reveals the pressure line by finding vectors with equal components in the second quadrant. The uni-directional structural change as the hydrostatic pressure increases is located in the second quadrant. (c) The contour map of calculated band gap values of varying $(\delta r, \delta c)$. Band gap reduction from $(\delta r, \delta c) = (0, 0)$ can be distinguishingly found in the second quadrant as well.

**Figure 4** Phonon frequencies are determined for the structures by changing $\delta c$ for the single Se chains. The highest mode for Se helix corresponds to a radial breathing mode. Phonon frequency suppression for the elongated chains is consistent with the Raman shift tendency observed in experiment. Se8 ring's phonon frequencies are indicated by the short bars.

fig1

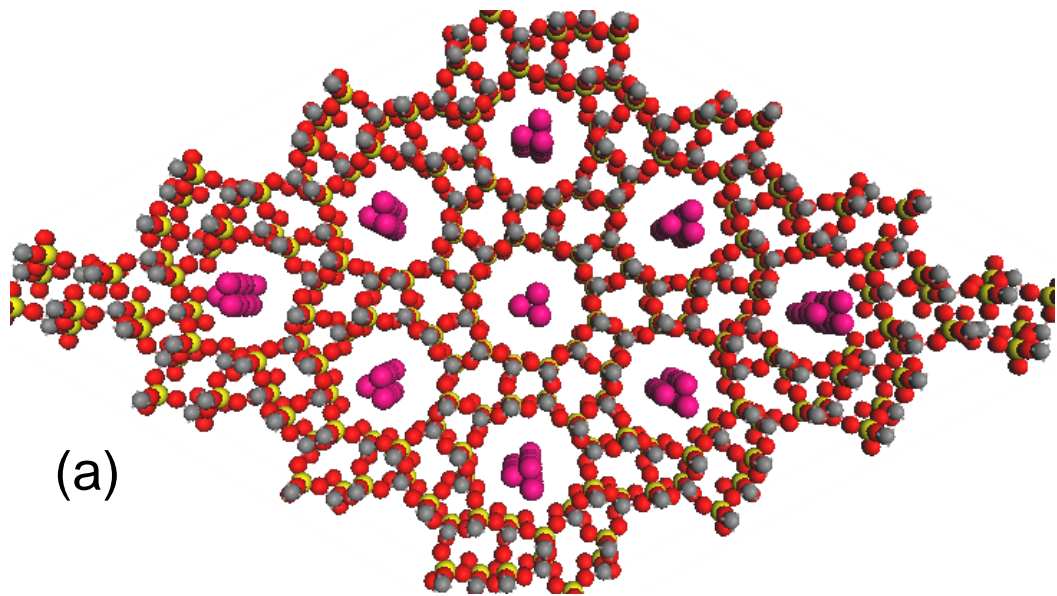

(a)

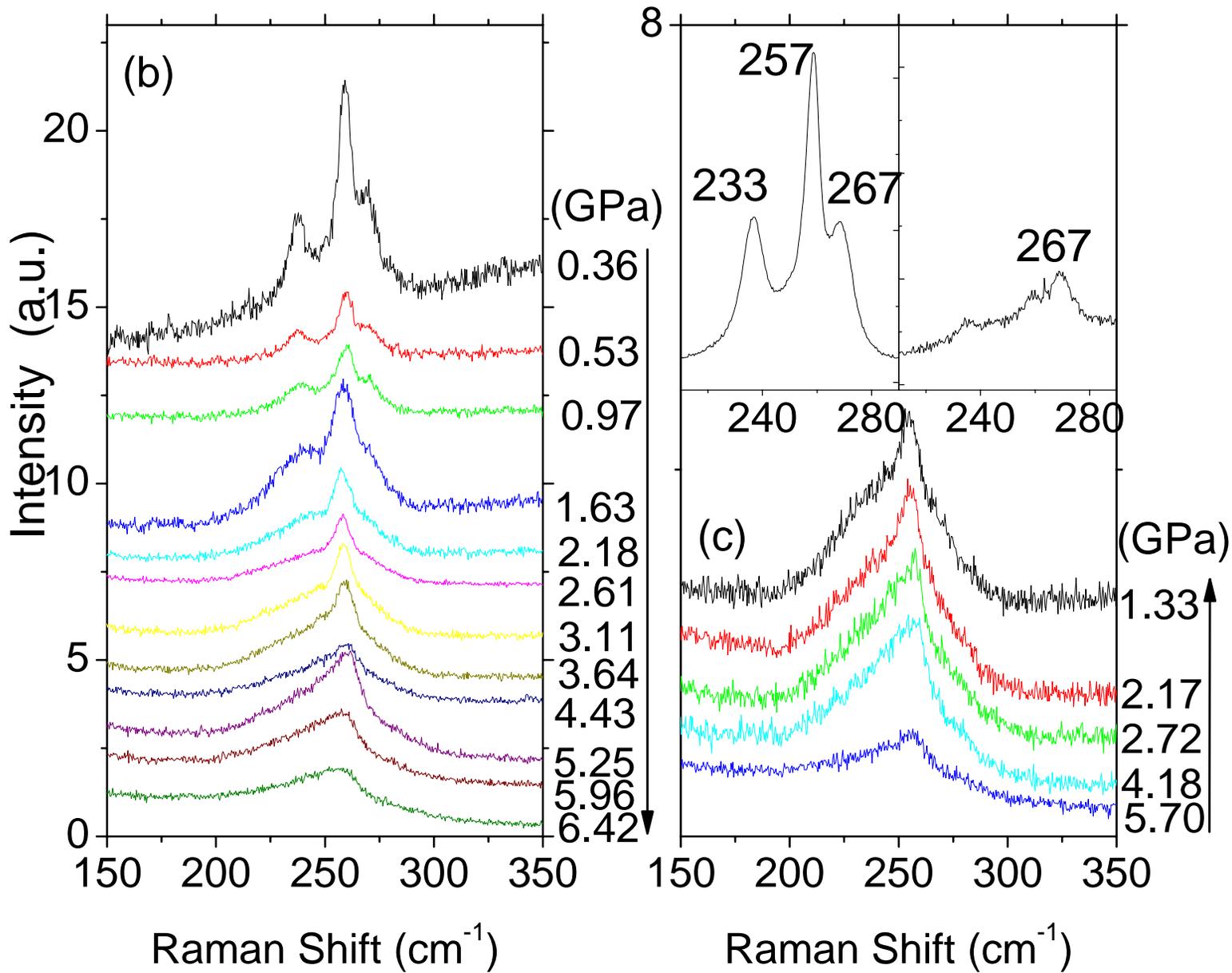

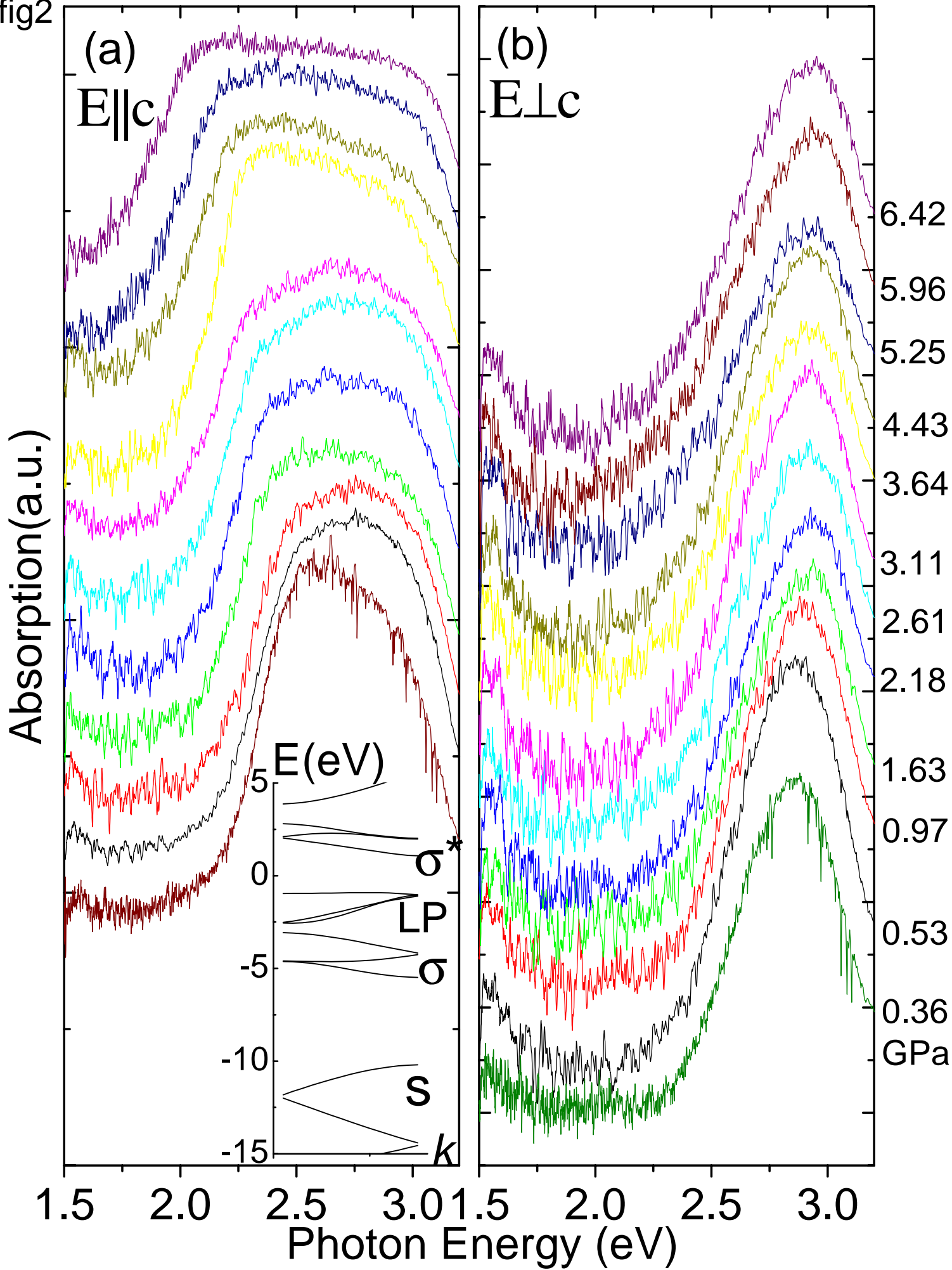

fig2

(a) E∥c

(b) E⊥c

6.42
5.96
5.25
4.43
3.64
3.11
2.61
2.18
1.63
0.97
0.53
0.36
GPa

E(eV)

0

σ*

LP

σ

-5

-10

S

-15

k

Absorption(a.u.)

Photon Energy (eV)

1.5    2.0    2.5    3.0    1.5    2.0    2.5    3.0

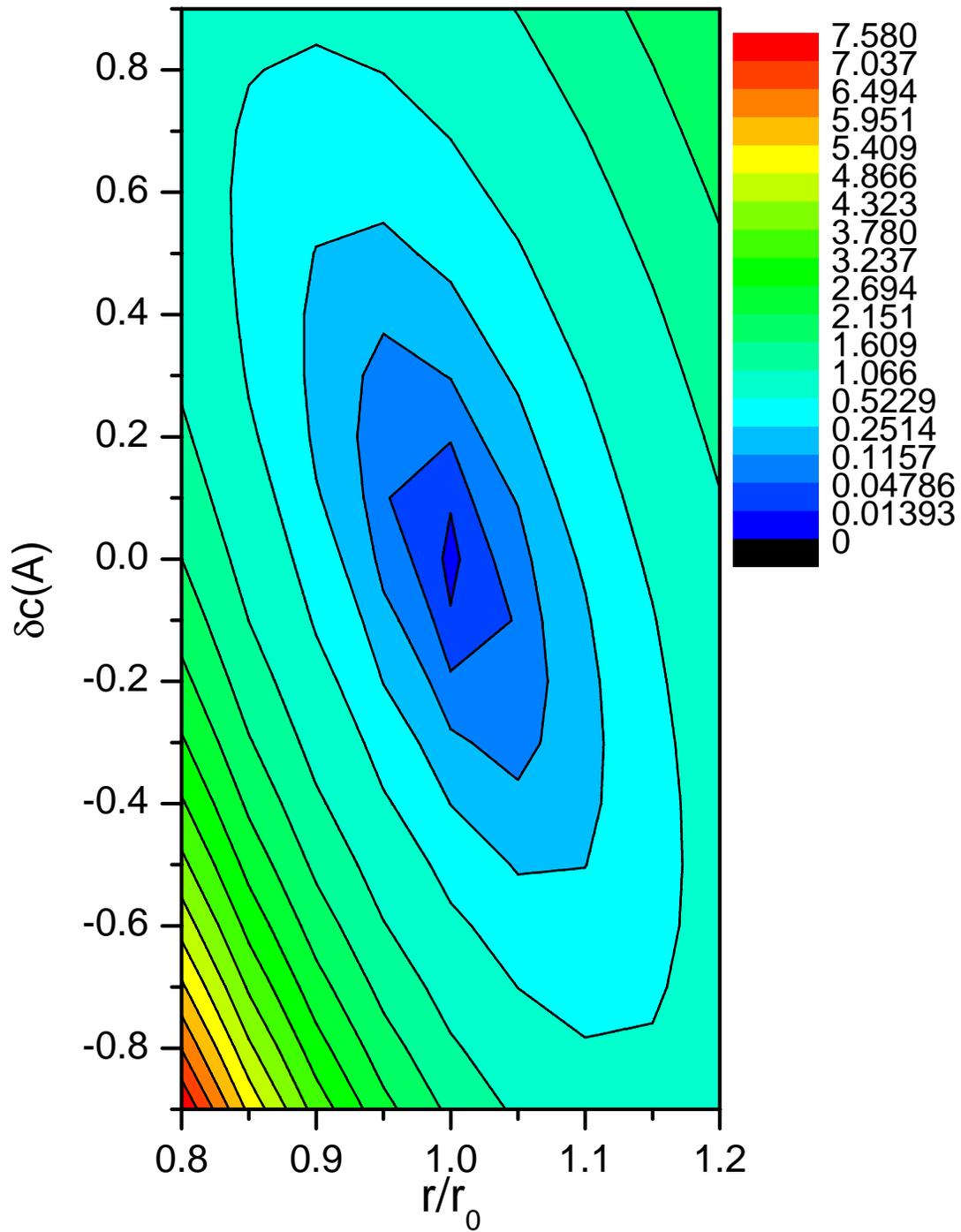

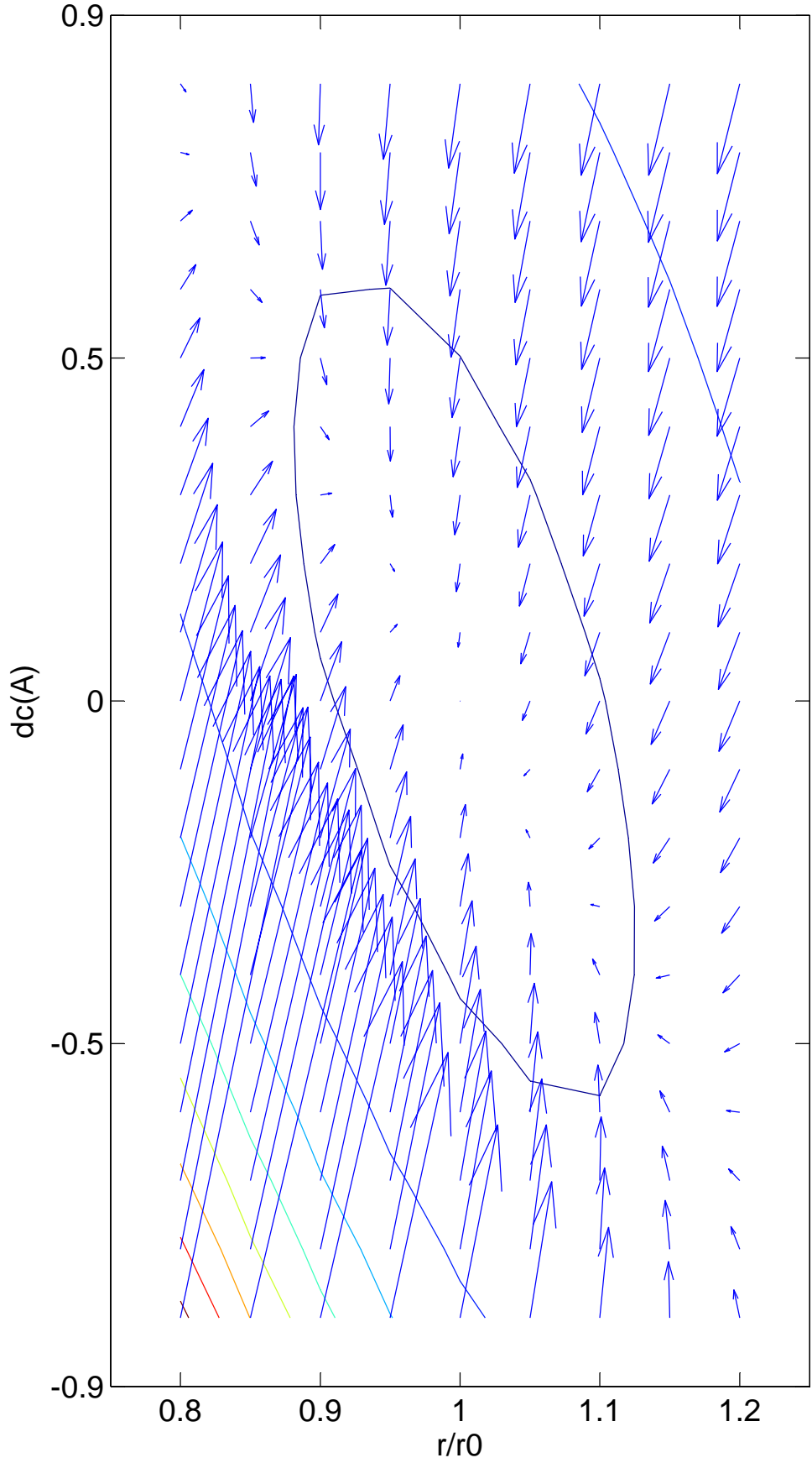

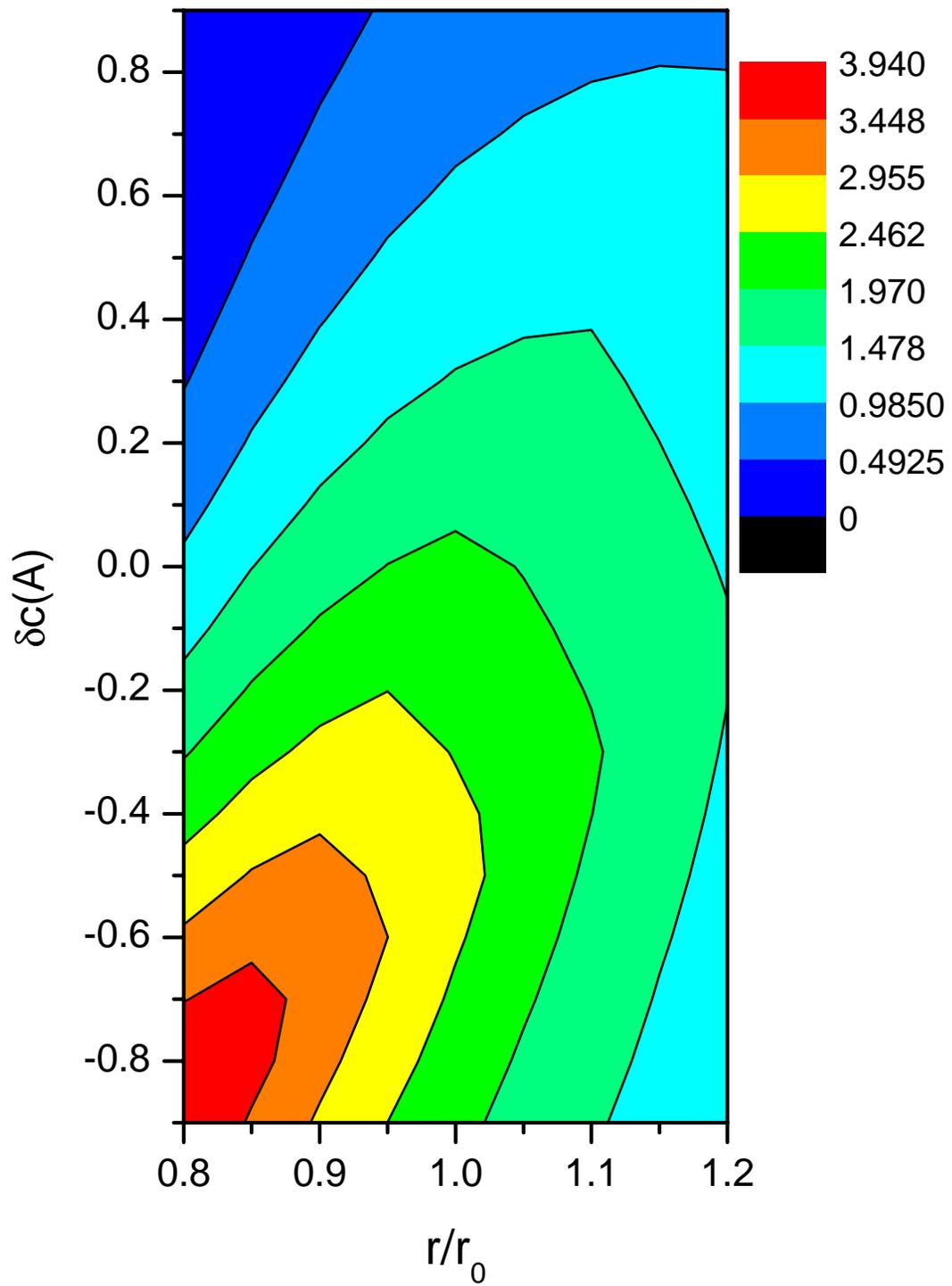

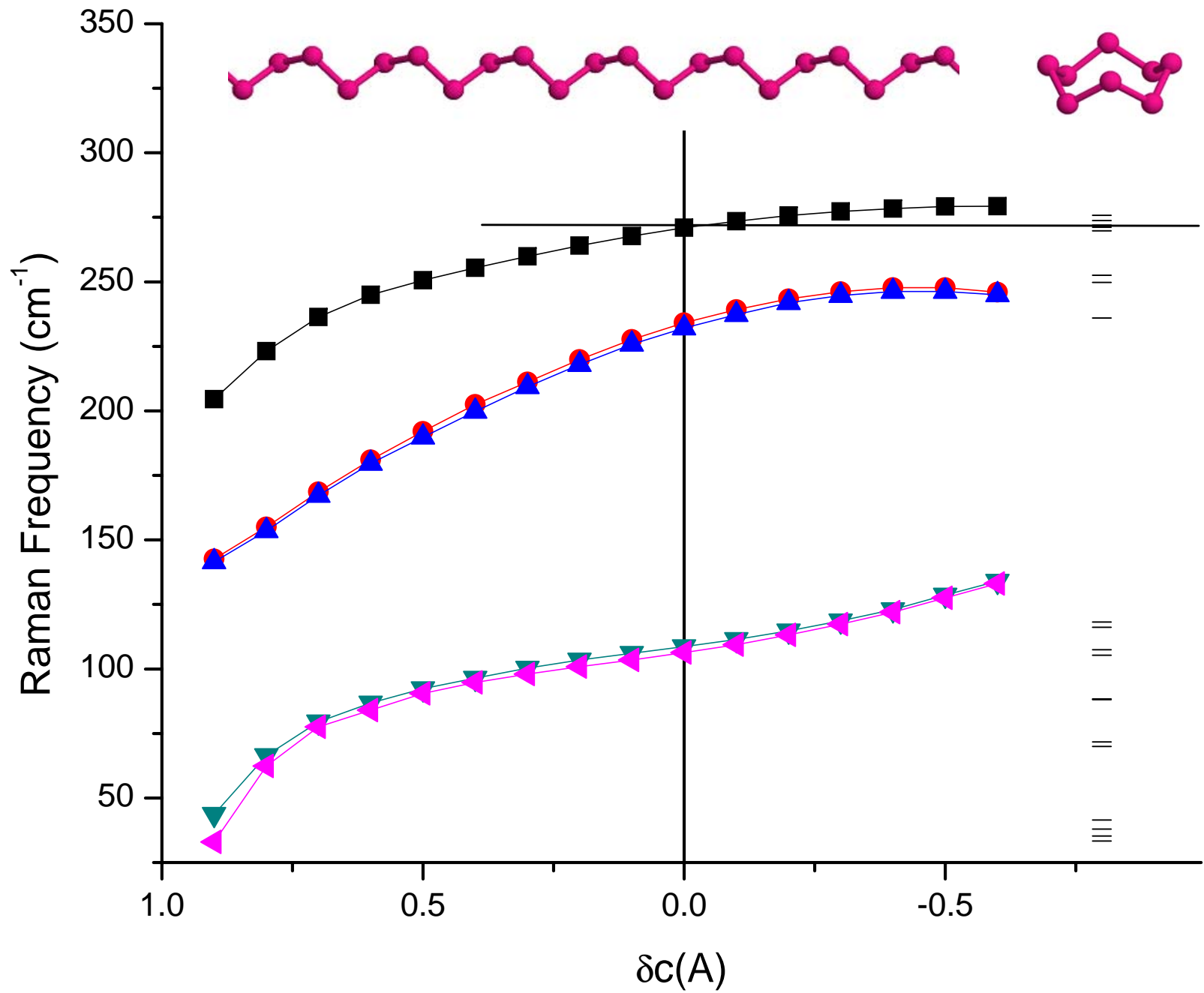

fig4